\documentclass[pra,twocolumn,titlepage,nofootinbib,amsmath,showpacs]{revtex4}%
\usepackage{amsmath}
\usepackage{amsfonts}
\usepackage{amssymb}
\usepackage{graphicx}%
\begin{document}
\title{Constraints on a long-range spin-independent interaction
from precision atomic physics}
\author{S.~G.~Karshenboim}
\email{savely.karshenboim@mpq.mpg.de} \affiliation{D.~I. Mendeleev
Institute for Metrology, St.Petersburg, 190005, Russia
\\ {\rm and}
Max-Planck-Institut f\"ur Quantenoptik, Garching,
85748, Germany}

\begin{abstract}
Constraints on a spin-independent interaction by exchange of a
neutral light boson are derived from precision data on the electron
anomalous magnetic moment and from atomic spectroscopy of hydrogen
and deuterium atoms. The mass range from $1\;$eV$/c^2$ to
$1\;$Mev$/c^2$ is studied and the effective coupling constant
$\alpha^\prime$ is allowed below the level of
$10^{-11}\!-\!10^{-13}$ depending on the value of the boson mass.
The mass range corresponds to the Yukawa radius from $0.0002\;$nm to
$20\;$nm, which covers the distances far above and far below the
Bohr radius of the hydrogen atom.
\end{abstract}
\maketitle

\section{Introduction}

Precision tests of quantum electrodynamics for bound states and free
particles allow us to verify various advanced methods of
measurements and calculations and determine precision values of
certain fundamental constants, such as the Rydberg constant
$R_\infty$ and fine structure constant $\alpha$. The overall
consistency of the results obtained for those constants~\cite{codata2006},
proves, in particular, that we can consistently
describe various fundamental phenomena in a broad range of distances
and energies~\cite{guide}.

Meanwhile, an introduction of new physics could affect
various scales with different strength and violate the consistency
mentioned. If such a problem is not observed, one sets various
constraints on possible new physics.

In particular, various unification theories suggest additional
particles  (see, e.g., \cite{weinberg,holdom,pospelov}). A specific
kind of such particles is a light neutral particle weakly
interacting with conventional matter consisting of electrons and
nucleons. Indeed, `lightness' in terms of particle physics ranges
from below 1 eV$/c^2$ to above 1 GeV$/c^2$. Stable neutral particles
of this kind can be also considered as a candidate for the dark
matter \cite{cosmo}.

In a broad range of distances one can consider an additional new
interaction as a modification of the Coulomb interaction such as
\begin{equation}\label{ar}
\frac{\alpha}{r} \to \frac{\alpha_{\rm eff}(r)}{r}= \frac{\alpha}{r}
+ \frac{\alpha^\prime} {r} e^{-\lambda r}\;.
\end{equation}
A similar substitute can be also written in the momentum space.

The particle is not necessarily stable in our consideration.
Generally, one should rather expect that a particle coupled to
charged particles, decays into a few photons, while a particle
coupled to leptons, could also decay to neutrinos. (Decay into a
pair of massive charged leptons can be forbidden for light particles
because of their lightness.) The substitute (\ref{ar}) is valid as
long as the decay width is much smaller than the mass $\lambda$.

Comparing values of the fine structure constant $\alpha$ obtained
from experiments with different characteristic distances and
momenta, one can check whether the Coulomb coupling constant is
really a constant. (We have in mind that the vacuum polarization
corrections responsible for the `running coupling constant' have
been already taken into account.)

The most precise measurements of various physical quantities are
often aiming at the determination of the values of fundamental
constants. The related precision data open a certain window of
opportunities to verify the constancy of the Coulomb coupling
constant. There are basically three important scales, which
contribute to precision QED-related experiments.

The scale of distances related to the Bohr radius
$a_0=1/(\alpha m_ec)$ is studied within hydrogen spectroscopy
with transitions involving the ground and the low excited states
($1s$, $2s$). The distances of about $10^3a_0$ are accessible
in experiments on high Rydberg states in the hydrogen atom
($n\simeq 30$), while the Compton wave length of the electron
$\lambdabar_C =\hbar/(m_ec)$ is
a characteristic distance involved in calculations of the electron
anomalous magnetic moment.

Altogether, the related distances vary from $2\times10^{-13}\;$m to
$2\times10^{-7}\;$m and related masses range from $1\;$eV$/c^2$ to
$1\;$MeV$/c^2$. Indeed, in the case of the completely relativistic
calculation of a correction to the electron anomalous magnetic
moment we cannot rely on a static component in (\ref{ar}), but
should deal with an interaction describing an exchange by an
intermediate particle.

Let us now consider precision QED-related experiments for these
three areas step by step.

\section{Anchor experiments: measurements on low states
in the hydrogen atom (distances $\sim (1-4)a_0$)}

A massive set of data on hydrogen and deuterium spectra is available
for low excited states related to physics from one to a few Bohr
radii. They are used here as an anchor for further comparison with
other distance/momentum scales. We remind that the Bohr radius
$a_0\simeq 0.53\times10^{-10}\;$m is a characteristic distance for
the electron orbit at the ground state. For the $2s$ state the
characteristic distance is $\sim 4a_0$.

Throughout the paper the relativistic units are applied in which
$\hbar=c=1$. In these units $a_0=1/\alpha m_e$. For the Yukawa
radius equal to the Bohr radius, the related mass of the
intermediate particle $\lambda$ is $3.5\;$keV, while the radius of
$4a_0$ corresponds to $1\;$keV.

To understand the procedure of evaluation, one can have in mind a
simple picture with the non-relativistic Schr\"odinger equation, because
all corrections beyond such approximation, which are due to QED and
relativistic effects, are well  under control and can be introduced when
necessary \cite{my_rep,EGS,codata2006}.

Spectroscopic data, which are the most important statistically, are
related to the $1s-2s$ transition \cite{mpq,mpq_d} and to the
$2s-ns/d$ transitions for $n=8,10,12$ \cite{paris}. We remind that
theory of hydrogen levels can be expressed in terms of the Rydberg
constant $R_\infty$ and the proton charge radius
\cite{my_rep,EGS,codata2006}. Those constants are not known from
other experiments with a sufficient accuracy and their best values
are determined from the spectroscopic data under question (see, e.g.,
\cite{codata2006}). Therefore, to separate these variables and to
determine $R_\infty$ one has to deal with at least two different
transitions \cite{lamb21,codata2006,my_rep}.

One of the utilized transitions is $1s-2s$ in hydrogen and
deuterium, for which the distances in the interval between $a_0$ and
$4a_0$ are involved. As to the other transitions, a certain
weighted average of data for all $2s-ns/d$ transitions is used. The
value of the principal quantum number $n$ for the involved excited
$ns/d$ states is substantially larger than 2, and because of that
the related Coulomb contribution is not larger than 1.5\% and can be
neglected as a good approximation.

So, the Rydberg constant \cite{codata2006}
\begin{equation}\label{ry12}
R_\infty=10\,973\,731.568\,527(73)\;{\rm m}^{-1}
\end{equation}
corresponds to $r\sim(1-4)a_0$. This value of $R_\infty$ can be
directly compared with another value related to a different distance
(see below for a comparison with a value related to $r\sim 10^3 a_0$).

Alternatively, one can extract a value of the fine structure
constant $\alpha$ from the result (\ref{ry12}) by applying the
relation
\begin{equation}
R_\infty=\frac{\alpha^2 m_ec}{2h}\;.
\end{equation}
For this purpose one should combine a value of $R_\infty$ with $h/M$
for a certain atom/particle and with a result or a chain of results,
which allows one to determine $M/m_e$ (see \cite{codata2006} for
detail).

The key point in such a determination of $\alpha$ is a measurement
of $h/M$. There are two high-precision independent experiments on that.
The first measurement is performed on caesium atoms
\cite{chu} and the most recent result leads to \cite{chu}
\begin{equation}
\alpha_{\rm Cs}^{-1}=137.036\,0000(11)\;.
\end{equation}
All auxiliary data are taken from \cite{codata2006}.

The other experiment has been performed on rubidium. The most recent
result is \cite{rb}
\begin{equation}
\alpha_{\rm Rb}^{-1}=137.035\,999\,45(62)\;.
\end{equation}

The weighted average of these two results\footnote{A somewhat less
accurate previous rubidium result $\alpha_{\rm
Rb}^{-1}=137.035\,998\,83(91)$ from the same group \cite{rbold} has
not been included into averaging because of possible correlations.
Once we treat that result as completely independent, the average
value is $\alpha^{-1}_{\rm atom}=137.035\,999\,39 (46)$ (cf.
(\ref{aa})), which does not change the final conclusions too much.}
\begin{equation}\label{aa}
\alpha^{-1}_{\rm atom}=137.035\,999\,59(53)
\end{equation}
is a value of the fine structure constant related to distances from
$a_0$ to $4a_0$.

\section{Precision measurements of highly excited
states in the hydrogen atom (distances $\sim10^3 a_0$)}

There is one more result on determination of the Rydberg constant,
which is usually not included into the adjustment \cite{codata2006}
because it is somewhat less accurate than the quoted above data and,
what is more important, the result is only a preliminary one. The
value
\begin{equation}\label{ry_mit}
R_\infty=10\,973\,731.568\,34 (69)\;{\rm m}^{-1}
\end{equation}
is obtained \cite{MIT} from a transition between the Rydberg states
with $n\simeq 30$ in the hydrogen atom. Since the value
resulted from a partial evaluation of possible systematic effects,
to be on the safe side the uncertainty in (\ref{ry_mit}) is tripled
against its original value. As it was confirmed by the authors of
the experiment \cite{kleppner}, that is rather an overconservative
estimation of the uncertainty.

Let us now constrain an effective interaction with mass $\lambda$,
which is
\begin{equation}
4\;{\rm eV}\ll\lambda\ll1\;{\rm keV} \;.
\end{equation}
A contribution of an additional boson in such a case reads
\begin{equation}\label{obs1}
 R_\infty=\left\{
 \begin{array}{ll}
 \frac{(\alpha+\alpha^\prime)^2m_e}{2}\;, ~~~& {\rm at} ~ r\sim a_0\;,\\
 ~\frac{\alpha^2m_e}{2}\;, & {\rm at}  ~r\sim 10^3 a_0\;.
 \end{array}\right.
 \end{equation}

That leads to a constraint \cite{prl}
\begin{equation}\label{const1}
\alpha^\prime=\Bigl(0.6\pm2.3\Bigr)\times 10^{-13}\;.
\end{equation}
That is the strongest constraint on a long-range spin-independent
interaction, which can be derived from atomic physics and
fundamental constants.

\section{Precision physics at the Compton wave length
(momentum $\sim m_ec$)}

One can also compare physics at the Bohr radius and at the Compton
wave length of an electron. A suitable tool for that is an
examination of various results on the fine structure constant. We
have already discussed a value related to $a_0$ (see (\ref{aa})).
Meanwhile, the most accurate value of the fine structure constant
comes from the anomalous magnetic moment of an electron
\begin{equation}
\alpha_{g\!-\!2}^{-1}=137.035\,999\,084(51)\;,
\end{equation}
obtained by combining the experimental result \cite{aexp} with
theory \cite{ath}.

Comparing it with the value in (\ref{aa}) one can find a constraint for
\begin{equation}
4\;{\rm keV}\ll\lambda\ll0.5\;{\rm MeV}\;.
\end{equation}
The results of the measurements can be presented as
\begin{equation}\label{obs2}
 \alpha=\left\{
 \begin{array}{ll}
 \alpha+\alpha^\prime\;, ~~~& {\rm at} ~ q\sim m_e\;,\\
 \alpha\;, & {\rm at}  ~q\sim 1/a_0\;,
 \end{array}\right.
 \end{equation}
where we refer to momentum space and assume that the intermediate
particle is a pure vector. The constraint reads  \cite{prl}
\begin{equation}\label{atomg2}
\alpha^\prime=\Bigl(2.7\pm2.9\Bigr)\times 10^{-11}\;.
\end{equation}

If the particle has another spin or it is a pseudovector or a
combination of vector and pseudovector (such as the $Z$ boson), then
the constraint is of the same order of magnitude but a factor
compatible with unity can appear.

\section{Final constraints}

\begin{figure}[thbp]
\begin{center}
\resizebox{1.0\columnwidth}{!}{\includegraphics{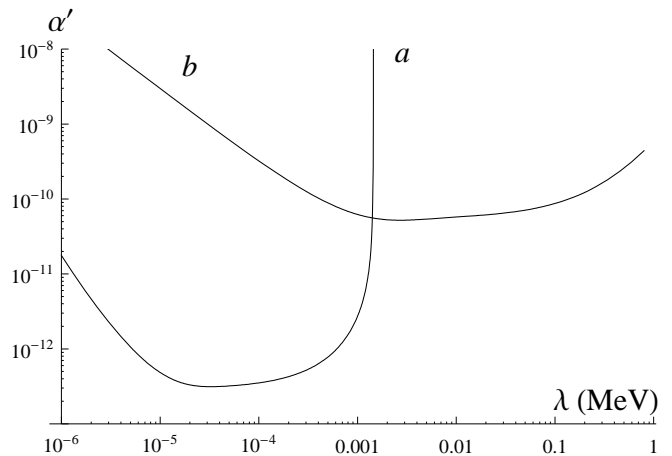}}
\end{center}
\caption{Constraints on a long-term spin-independent interaction
from hydrogen spectroscopy and $g_e-2$, including a constraint from
hydrogen spectroscopy with low and Rydberg states ($a$), a
comparison of low states and $g_e-2$ ($b$). The lines are for
$|\alpha^\prime|$ and the confidence level corresponds to one
standard deviation.}
\label{f:const}       
\end{figure}

Our results are summarized in Table~\ref{t:sum}. We can extend our
analysis beyond the asymptotic limits. In particular, we also present in
Fig.~\ref{f:const} results, which cover an intermediate area of a few
keV \cite{prl}.

\begin{table}[htbp]
 \begin{center}
 \begin{tabular}{ccc}
 \hline
Range  & $\alpha(r)-\alpha (a_0)$ & Effect \\[0.8ex]
 \hline
$4\;{\rm eV}\ll\lambda\ll1\;{\rm keV}$
&$\bigl(-0.6\pm2.3\bigr)\times 10^{-13}$& H, $n=30$ \\[0.8ex]
$4\;{\rm keV}\ll\lambda\ll0.5\;{\rm MeV}$&$\bigl(2.7\pm2.9\bigr)\times 10^{-11}$& $g_e-2$\\[0.8ex]
\hline
 \end{tabular}
\caption{The constraint on the deviation of the effective long-range
interaction $\alpha(r)/r$ from the Coulomb exchange \cite{prl}. Here,
$\alpha(r)=\alpha(\infty)+\alpha^\prime \exp(-\lambda r)$. The $a_0$
scale is related to hydrogen spectroscopy for transitions involving
low states ($1s$, $2s$). The related distance range is
$r=0.5\times10^{-7}\;$m (for $\lambda=4\;$eV),
$0.5\times10^{-10}\;$m (for $\lambda=4\;$keV) and
$0.4\times10^{-12}\;$m (for $\lambda=0.5\;$MeV).
 \label{t:sum}}
 \end{center}
 \end{table}

To extend the constraint to a region $\lambda \sim 1/a_0$ and
$\lambda\sim m_e$, we have to modify observational equations
(\ref{obs1}) and (\ref{obs2}).

Instead of (\ref{obs1}) we find
\begin{equation}\label{obs11}
 R_\infty=\left\{
 \begin{array}{ll}
 \frac{\bigl(\alpha+\alpha^\prime{\cal F}_{12}(\lambda/(\alpha m_e))\bigl)^2m_e}{2}\;, ~~~& {\rm at} ~ r\sim a_0\;,\\
 ~\frac{\bigl(\alpha+\alpha^\prime{\cal F}_{30}(\lambda/(\alpha m_e))\bigl)^2m_e}{2}\;, & {\rm at}  ~r\sim 10^3
 a_0\;,
 \end{array}\right.
 \end{equation}
where
\begin{eqnarray}
{\cal F}_{12}(x)&=& 4\left[\left(\frac{1}{1 +x}\right)^2-2
\left(\frac{1}{1+x}\right)^3+\frac{3}{2}
\left(\frac{1}{1+x}\right)^4\right]\nonumber\\
&&-\left(\frac{2}{2+x}\right)^2\;,
\end{eqnarray}
and for ${\cal F}_{30}(\lambda/(\alpha m_e))$ we have to consider
a function
\begin{equation}
{\cal
F}_{30}(x)=\frac{\frac{1}{(n+1)^2}\left(\frac{2}{2+(n+1)\,x}\right)^{2n+1}-\frac{1}{n^2}
\left(\frac{2}{2+n\,x}\right)^{2n-1}}{1/(n+1)^2-1/n^2}\;,
\end{equation}
related to $n\simeq 30$. The transitions studied in \cite{MIT} were
transitions between the circular states ($l=n-1$): $27\to 28$ and
$29\to30$.

The related constraint is
\begin{equation}\label{const11}
\alpha^\prime=\frac{\bigl(0.6\pm2.3\bigr)\times 10^{-13}}{
 {\cal F}_{12}(\lambda/(\alpha m_e))- {\cal F}_{30}(\lambda/(\alpha m_e))}\;.
\end{equation}

Eq.~(\ref{obs2}) is to be substituted for
\begin{equation}\label{obs21}
 \alpha=\left\{
 \begin{array}{ll}
 \alpha+\alpha^\prime{\cal F}_{g\!-\!2}(\lambda/m_e)\;, ~~~& {\rm at} ~ q\sim m_e\;,\\
 \alpha+\alpha^\prime{\cal F}_{12}(\lambda/(\alpha m_e))\;, & {\rm at}  ~q\sim 1/a_0\;,
 \end{array}\right.
 \end{equation}
where the function \cite{kaquote}
\begin{eqnarray}
{\cal F}_{g\!-\!2}(x)&=&  2\Bigl[-\bigl(x^5 - 4x^3 + 2x \bigr) \;
\frac{\tan^{-1}\sqrt{ \frac{4-x^2}{x^2}
}}{\sqrt{4-x^2}}\nonumber\\&& + \bigl(x^4 - 2 x^2 \bigr) \ln{x} -x^2
+ \frac{1}{2}\Bigr]\nonumber
\end{eqnarray}
is the function which is applied when one calculates the
hadronic vacuum polarization through integration over a dispersion
variable $\lambda=\sqrt{s}$.

The related constraint is
\begin{equation}\label{const21}
\alpha^\prime=\frac{\bigl(2.7\pm2.9\bigr)\times 10^{-11}}{{\cal
F}_{g\!-\!2}(\lambda/m_e)-{\cal F}_{12}(\lambda/(\alpha m_e))}\;.
\end{equation}

\begin{figure}[thbp]
\begin{center}
\resizebox{0.8\columnwidth}{!}{\includegraphics{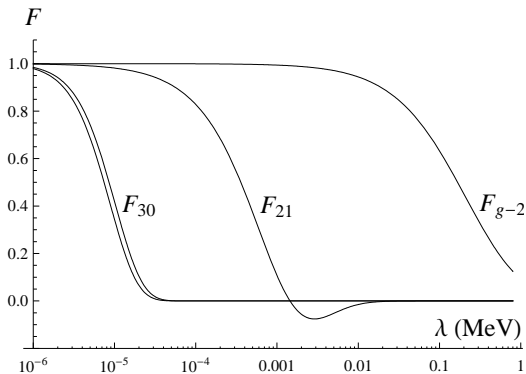}}
\end{center}
\caption{The profile functions, which determine coverage of the
constraints on a long-term spin-independent interaction from
hydrogen spectroscopy and the anomalous magnetic moment of an
electron. The profile function ${\cal F}_{30}(x)$ is with $n=27$
 and $n=29$.}
\label{f:2729}       
\end{figure}

We summarize behavior of the profile functions ${\cal F}$ in
Fig.~\ref{f:2729}. Each of them is equal to unity for $\lambda\to 0$
and equal to zero for $\lambda\to \infty$. The unity region is the one,
where the Yukawa potential is indistinguishable from the Coulomb
potential, and the effective Coulomb coupling constant there is
$\alpha+\alpha^\prime$. In the zero region the Yukawa potential
shrinks to a zero-radius potential and does not affect any
measurable values.

What is different in the profile functions is the characteristic
value of $\lambda$, which separates the unity region from the zero
one and the shape of the transition area around it.

We present in Fig.~\ref{f:2729} four rather than three profile
functions. Function ${\cal F}_{30}$ is presented in two versions. That
is due to the fact that the result (\ref{ry_mit}) is a preliminary
one and the evaluation has not yet been completed. The data included
two transitions, namely $27\to 28$ and $29\to30$ and it is
in part uncertain what their relative weights should be in
averaging. To clarify the issue, we present in Fig.~\ref{f:2729} two
versions of the profile functions ${\cal F}_{30}$, namely for $n=27$ and
$n=29$. One sees that the results are nearly identical in the logarithmic
scale. Due to that, for the final constraint in Fig.~\ref{f:const}
we have applied a profile with $n=28$.

\section{Summary}

The interaction with an intermediate particle is not necessarily
universal. The hydrogen and deuterium data allow one to conclude about
$\alpha_{ep}$ and $\alpha_{ed}$, while the anomalous magnetic moment
deals with $\alpha_{ee}$. (The uncertainty of the anchor data
involving hydrogen and deuterium is small and the accuracy of
the constraint is determined in both cases above by the other pieces
of data. That allows to reduce the $a_0$ related data to pure
hydrogen data.) A possible discrepancy between these two values
could be interpreted not only as related to the certain distance
scale, but also to non-universality of one-particle exchange within
a broader range.

We remind that the neutrality of a hydrogen atom or a neutron (see,
e.g., \cite{pdg}) is a property related to a macroscopic distance
and, in principle, a direct test whether there is an ultraweak
long-range component of interaction of a hydrogen atom or a neutron
at microscopic distances is required. We will address this problem
elsewhere.

\begin{figure}[thbp]
\begin{center}
\resizebox{1.08\columnwidth}{!}{\includegraphics{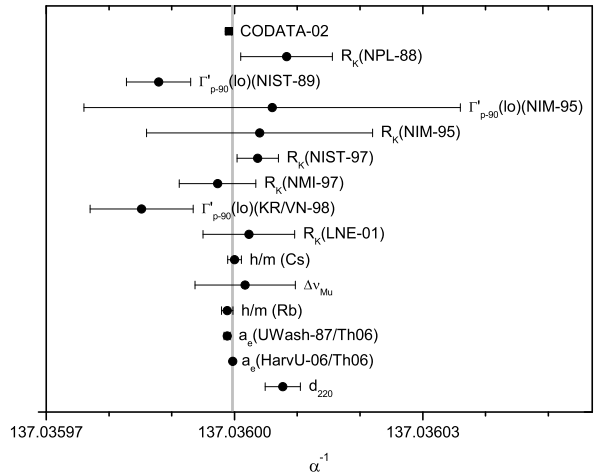}}
\end{center}
\caption{Determination of the fine structure constant $\alpha$ in
the CODATA-2006 adjustment \cite{codata2006}, where the references
can be found. The vertical band stands for the adjusted CODATA-2006
value. The figure is reproduced from \cite{fig_ref}.}
\label{f:alp}       
\end{figure}

For the $b$ constraint in Fig.~\ref{f:const}, we have utilized two
values of the fine structure constant, namely those from $g_e-2$ and
from atomic physics. In principle, one can use other less accurate
values (see Fig.~\ref{f:alp} and \cite{codata2006} for a review of
data and references) to compare with $\alpha_{g\!-\!2}$. That
produces a constraint similar to the $b$ line in Fig.~\ref{f:const}. It
is weaker for $\lambda \gg a_0^{-1}$, but has the same profile
behavior, while for $\lambda \leq a_0^{-1}$ the limitation increases
slower than the $b$ line. However, as we see in Fig.~\ref{f:const}, in
this area the $a$ constraint is already applicable and such
extension of the $b$ constraints to the lower mass range cannot
improve the whole constraint.

Another issue is model dependence. The substitute (\ref{ar}) is to
constrain a situation with a single light intermediate particle. If
there are a few of them and their coupling constants are comparable,
the situation becomes more complicated. Still, if the related
coupling constants are of the same sign, the results of the paper
are still applicable. However, if the Yukawa terms involve
interactions with different radii, while the coupling constants have
compatible absolute values and opposite signs, the constraint
derived is completely out of validity. As an example, one can
consider a long-range correction of the form
\[
\alpha^\prime\frac{e^{-\lambda_1 r} - e^{-\lambda_1 r}} {r}\;,
\]
where the masses are of the same order (e.g.,
$\lambda_2=2\lambda_1$). Indeed, in this case the sensitivity of
atomic energy levels to $\alpha^\prime$ is substantially reduced in
most of the area, where the constraint in Fig.~\ref{f:const} is
effective.

Concluding we emphasize that  atomic physics allows one to access
distances much larger than the Bohr radius and our limitation on
$\alpha^\prime$ is valid for distances up to $0.02\;\mu$m. That can
be compared with the limitation on the fifth force from
Casimir-effect experiments \cite{casimir}, which also reached the
range below $1\;\mu$m. Comparing the strength, we note that in the
case of Casimir effect the conventional parameterization for the
Yukawa potential is $G m_1 m_2 \tilde{\alpha} e^{-
r/\tilde{\lambda}}/r$, where $m_i$ is a mass of a bulk body and $G$
is Newtonian gravitation constant. There is no direct
model-independent correspondence between mass-related and
charge-related parameterization. The most important part of the
conversion factor is $G m_p^2\sim 6\times10^{-39}$ (since in bulk
matter there is roughly 0.5-1 charge, e.g., baryon or lepton charge,
per nucleon--the rough factor of 0.5 comes from the fact that we
have an electron per from 2 to 2.5 nucleons). Using it, we find that
our constraints starting from a distance below 30 nm, where the
Casimir-effect constraints stop, are weaker by a few orders of
magnitude. However, they are extended to shorter distances, not
accessible for the Casimir-force related experiments.

What is more important, our constraint is complementary to the
Casimir-effect limitation. The latter deals with the neutral
bulk matter only and cannot constrain a massive ultraweak photon
$\gamma^*$.  In contrast to that, our constraint in
Table~\ref{t:sum} and in Fig.~\ref{f:const} is covering such a case.
Our results for the coupling constant $\alpha^\prime$ related to
$\gamma^*$ are consistent with zero and the limitation at atomic
distances is in the range of a few parts in $10^{13}$ to a few parts in
$10^{11}$ depending on details as discussed above.

The keV mass range has been partly explored by means of astrophysics
and cosmology. Such constraints \cite{astro,cosmo} mostly deal with
real particles and their propagation through the matter. Details,
such as the lifetime, various couplings and collision rates, are
involved and they may be related to results in terms of
$\alpha^\prime$ and $\lambda$ only through certain model-dependent
relationships. Our constraints, free of such model-dependent
suggestions, are complementary to those.

Another option, also not covered by Casimir effect is due to a
spin-dependent long-range interaction to be considered elsewhere.

\section*{Acknowledgements}

This work was supported in part by RFBR (grants \#\# 08-02-91969 \&
08-02-13516) and DFG (grant GZ 436 RUS 113/769/0-3). The author is
grateful to Dan Kleppner, Andrej Afanasev, Vladimir Korobov, Masaki
Hori, Astrid Lambrecht, Dmitry Toporkov, Eugene Korzinin, Simon
Eidelman, and  Maxim Pospelov for useful and stimulating
discussions.

\end{document}